# Inventions on Adaptable Menu
## A TRIZ based analysis

**Umakant Mishra**

Bangalore, India

http://umakantm.blogspot.in

**Contents**

## 1. Introduction

The menu is one of the most widely used elements of a graphical user interface. The objective of a menu system is to provide various commands and functions to the user in an easy way so that the user can just select the desired operation from a given list instead of typing a complex command in the command prompt.

There are several aspects of a menu system such as functionality, adaptability, aesthetics, presentation, navigation etc. Each of these factors is important to make a menu system powerful and effective. This article will discuss about the flexibility and adaptability of the menu system. Flexibility and adaptability is one of the desirable features of an advanced menu system.
.

## 1.1 Need for adaptability

In a conventional menu system the menu items or options are hard-coded in the computer program. The programmer or developer composes menu items at the time of development. The developer tries to include all options that he feels may be required by the user in future. Although the items are decided from "requirement analysis" and other studies, it is difficult to know the exact need of a user at a future period of time. This leads to inclusion of a lot of items in the menu, which leads to user confusion and frustration.

If the developer reduces the number of menu items, there is a greater chance for the user not finding a desired option. If the developer adds all the important items to the menu, the user is confused and frustrated to see a long list of items. This creates a contradiction as the menu should contain maximum number of items to satisfy user requirement and at the same time should contain the minimum number of items to reduce user confusion and frustration.

Thus there is a need for adaptable menu that can be changed according to user requirement. The items should change from user to user and from time to time.

## 1.2 The mechanism of adaptability

As we saw above, it is not possible for the developer to add the exact number of menus that a user may need in future. This is not only because it is difficult to predict the future needs but also because the requirement varies for different users at different times. The problem can be addressed by an adaptable menu that can change according to the user requirement. The customization of the menu can be done either manually by the user or automatically by the system.

In a manual customization system, the menu details are kept in a separate database and not hard coded in the main program. The user is given a separate menu-editing interface to add or remove any item he needs in the item list. The menu engine accesses the menu items from the menu database and displays the desired items to the user. The user changes the options as and when there is a need.

In an automatic system, the items are dynamically changed by the menu system, based on the usage pattern of the user. In general, more frequently accessed items are given a higher level of priority. The adaptable system organizes the items in various ways based on their priority levels. The high priority items may be displayed on the top of the list for easier access. Alternatively the system may display only the high priority items by default and display low priority items on the event of a special request by the user.

## 2. Using TRIZ to build adaptable menu

The ideality, contradictions and Inventive Principles can be used to solve the issues related to an adaptable menu. Assuming that the readers are familiar with TRIZ methods and principles I may be excused for directly getting into analysis without giving any introduction to TRIZ and its methods.

**2.1 The Ideal Final Result (IFR)**

A menu system tries to achieve several objectives from different perspectives. One of the important IFR of the menu system is the following.

- The menu should automatically display only the desired item to the user (this would maximize the ease of understanding, navigation, selection, space utilization and other benefits).

But this is a hard to achieve IFR. If the system can display only the desired item then the system can also select that item for the user thereby eliminating the need for displaying a menu at all. Let's look at the IFR relating to the flexibility and adaptability of the menu.

- The system should organize and display the most desirable options automatically to the user.

- The user should also be able to do any modifications (add, delete, modify, reorganize etc.) to the menu items without acquiring special skills.

- The menu should be capable of displaying any number of items that may be required by a user without blocking the information on the screen.

**2.2 Contradictions in achieving IFR**

We face the following contradictions while trying to achieve the above IFR for an adaptable menu.

**Contradiction**: The menu should display maximum number of items to include all the items the user may need. The menu should display minimum number of items to exclude all the items the user may not need.

**Solution**: Allow user customization.

**Contradiction**: The user should be able to customize the menu items and characteristics as may be useful for him. But the user need not learn the programming language or recompile the program.

**Solution**: Provide a customization interface so that the user can customize the menu without needing to change the source code.

**Contradiction**: The menu should be customized for the user, but the user should not spend time and effort for customization.

**Solution**: Allow automated customization according to usage pattern. The user may modify only in exceptional cases.

## 3. Inventions on adaptable menu

There are several inventions on flexibility and adaptability of a menu system. The following are some of the selected cases from US patent database.

### 3.1 Methods and apparatus for dynamic menu generation in a menu driven computer system (5041967)

**Background problem**

A conventional menu system has rigidly defined menu items, which are hard-coded in the menu program. This feature is not very convenient, as it cannot cater to the varied requirements of different users. The user cannot change the items, as they are hard coded inside the program. There is a need to allow the user to change the menu as per his/ her need.

**Solution provided by the Invention:**

Patent 5041967 (Invented by Ephrath et al., Assigned to Bell Communication Research Inc, Aug 91) discloses a method of user defined, dynamically generated multilevel menus. According to the invention, the definitions will be stored in a menu table, which can be edited by the user as per the local need.

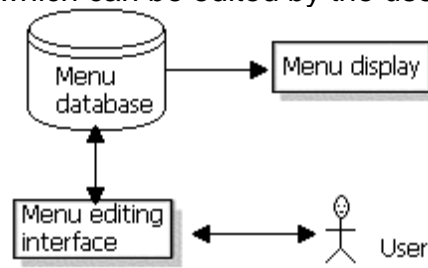

This solution is very simple but nevertheless a powerful and fundamental solution.

**TRIZ based analysis**

The user should be able to customize the menu items to cater to his need, as the user knows his need. But the user has no access to the source code to modify the menu. The developer can modify menu, but he does not know the varying user needs at a future period of time (Contradiction).

The invention separates the menu items from the menu engine and stores in a separate database (Principle-2: Taking out).

The user is allowed to edit the items as required (Principle-15: Dynamize).

### 3.2 Method and apparatus for selecting button functions and retaining selected options on a display (5243697)

**Background problem**

A menu is very important in a graphical user interface as it can provide a large number of options to the user without occupying any permanent screen space. But the major drawback of the menu mechanism is that the user has to open the menu again and again each time he wants to select a particular item. There is a need to keep the frequently accessed menu items displayed on the screen so that the user can access the desired item without opening and navigating the menu every time.

**Solution provided by the invention**

Patent 5243697 (invented by Hoeber et al., assigned by Sun Microsystems, issued Sep 1993) discloses a method of using a pushpin to retain the frequently used menus on the display screen.

According to the invention when the menu button is selected by the user, a menu appears in a rectangular box containing a pushpin and several menu items. If the user clicks on the pushpin then the temporary menu box is converted into a permanent window which remains on the display regardless of other display operations. The user may click again on the pushpin to release the permanent placement of the menu on the screen.

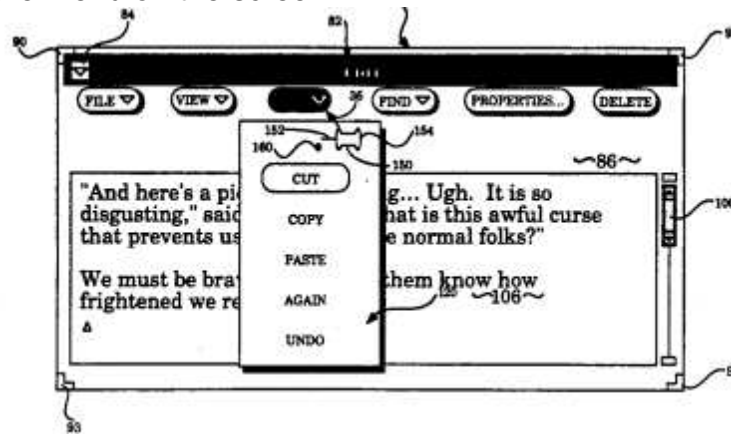

This invention provides an intuitive means for retaining frequently used menus on a graphical user interface screen.

### TRIZ based analysis

The invention keeps the menu screen opened for successive use, instead of needing the user to open it again and again every time he needs (Principle-10: Prior action).

The user can "pin" the menu to keep it open or "release the pin" for normal use (prnicple-15: Dynamize).

According to the invention, when the pushpin is clicked, the icon of the pin is modified so that it appears that the pushpin has "pinned" the menu to the screen (Principle-35: Parameter change, Principle-23: Feedback.)

### 3.3 Menu editor for a graphical user interface (5760776)

### Background problem

The menu editor is used to define the properties of an individual menu or submenu. The menu editors in the prior art typically display only one menu for editing. Besides while selecting the associated functions for the menu items, if the user clicks wrongly on the associated function of the menu item that may execute the function or may lead to an undesirable result. There is a need for an improved visualization and safer means of editing the menus.

### Solution provided by the invention

Patent 5760776 (invented by McGurrin et al., assignee Oracle Corporation, issued Jun 1998) disclosed a method of improved visualization and safer manipulation of menu structure during the editing process. The menu editor displays multiple branches of menus to edit simultaneously.

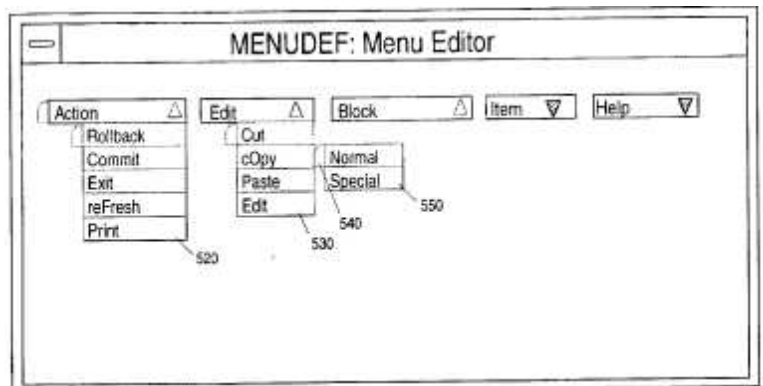

This menu editor is useful to create custom menus for user's applications.

### TRIZ based analysis

The invention displays multiple branches of menu to edit simultaneously (Principle-7: Nested doll).

The new method does not select the function of the menu by clicking on that option but by "grabbing" the handle to that function. This reduces the accidental execution of unintended actions (Principle-9: Prior counteraction).

### 3.4 Method and apparatus for selectively expandable menus (5801703)

### Background

Some applications have large number of options in the menu which make it too complex for the user to navigate. Besides, the user has to navigate through a number of sub-menus each time a feature is desired. How to simplify the access?

One solution is to use keyboard short cuts instead of menu items. But the key combinations are difficult to remember. The other solution is a "tear off" menu that locks a particular menu/submenu open for future selection. But there is no easy way to customize a "tear off" menu.

### Solution provided by the invention

Bowden et al. invented a menu (Patent 5801703, assigned by Island Graphics Corporation, issued in Sep 1998) that is selectively expandable.

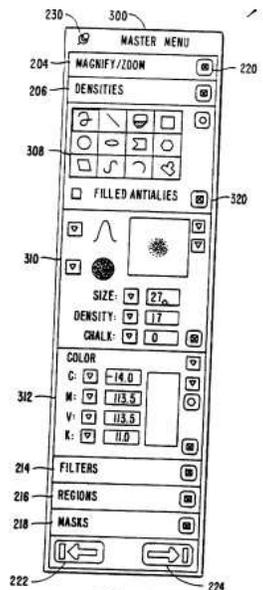

The invention displays the master menu, which contains several items and panels. Each panel is made in such a way that it can be expanded or contracted. The user expands the panel and makes selections from the expanded menu panel without letting it automatically contracted.

The user is able to expand as many menu panels as desired and able to customize the master menu.

### TRIZ based analysis

The invention keeps the sub-menus or panels expanded selectively which are frequently used (Principle-10: Prior Action).

There is provision for the user to keep the panels expanded or contracted as required (Principle-15: Dynamize).

The user can expand multiple menu panels (Principle-40: Composite).

### 3.5 Auto adjustable menu (6121968)

**Background problem**

The designer of computer applications provides more and more commands in the menus for easy access to the user. However, it suffers from its negative side that the menus become cluttered and look confusing to the users.

**Solution provided by the invention**

US Patent 6121968 (Invented by Arcuri et al, assigned to Microsoft, Sep 2000) discloses a method of dynamically changing the available commands in a given short menu based upon the usage pattern of the user. Initially the menu opens in a short menu mode. The short menu can be dynamically expanded to long menu mode to display the complete set of commands.

The items in the short menu changes according to the usage pattern. The most frequently used commands are added to the menu and less frequently used items are removed. This adaptive menu solves most of the earlier problems.

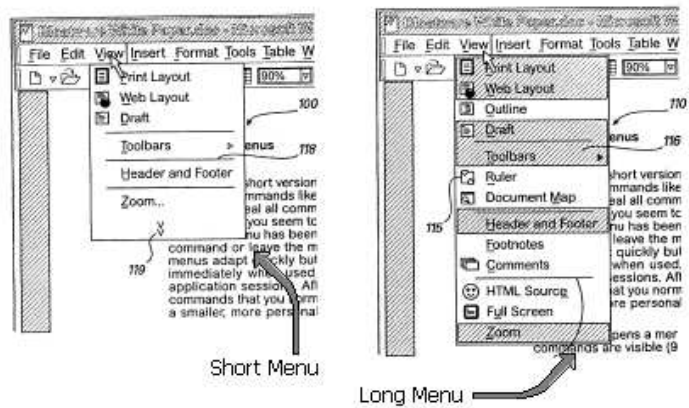

**TRIZ based analysis**

The menu should contain enough of items/ commands to give maximum options to the user. At the same time, it should contain minimum number of items to consume less space and facilitate easy navigation (Contradiction).

The invention adjusts the size of menu to be short or long dynamically (Principle-15: Dynamize).

The items of the menu list change dynamically according to the usage pattern of the user (Principle-15: Dynamize).

More important items are automatically added to the list and less important items are automatically removed from the list (Principle-25: Self Service).

### 3.6 Using heuristic factors to manage the dynamic changes in the menu options (6266060)

### Background

The prior art menu management systems consider only one or two factors to organize the menu items. Besides it does not allow a user to choose which heuristic factors are considered to tailor the menu items. There is a need to consider multiple heuristic factors to control an improved menu management mechanism.

### Solution provided by the invention

Steven Roth invented a method (Patent 6266060, assigned to IBM, July 01) of using heuristic factors to manage the dynamic changes in the menu options. The term heuristic is used to generally describe information about past use.

The invention provides several discrete yet complementary features for menu management. One feature is automatic menu arrangement based on a combination of frequency and recency of selection. Another feature is the consideration of time of the day for menu arrangement. This allows the user to order that certain menu items are presented during certain hours of the day.

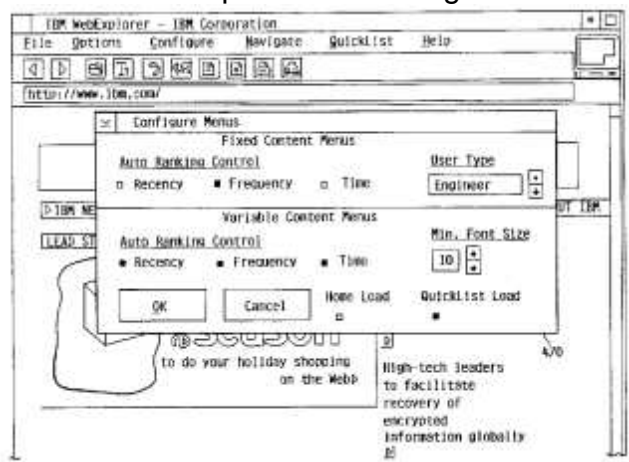

**TRIZ based analysis**

The invention uses multiple heuristic factors such as frequency of selection, recency of selection, time of selection etc. to arrange the items for user convenience. (Principle-40: Composite).

## 4. Summary and conclusion

All the above inventions try to achieve the Ideal Final Result and try to solve the associated contradictions in some or other way. Some of them provide very powerful solutions. The older inventions provide more manual customization of menu whereas the later inventions provide more automated customization. The future inventions will present even better solutions to achieve the same IFR.

## Reference:

1. US Patent 5041967, "Methods and apparatus for dynamic menu generation in a menu driven computer system", Invented by Ephrath et al., Assigned to Bell Communication Research Inc, Aug 91

2. US Patent 5243697, "Method and apparatus for selecting button functions and retaining selected options on a display", invented by Hoeber et al., assigned by Sun Microsystems, issued Sep 1993.

3. US Patent 5760776, "Menu editor for a graphical user interface", invented by McGurrin et al., assignee Oracle Corporation, issued Jun 1998.

4. US Patent 5801703, "Method and apparatus for selectably expandable menus", Invented by Bowden et al., Assigned to Island Graphics Corporation, Sep 98.

5. US Patent 6121968, "Adaptive Menus", Invented by Arcuri et al, assigned to Microsoft, Sep 2000

6. US Patent 6266060, "Menu management mechanism that displays menu items based on multiple heuristic factors" Invented by Roth, Assigned to IBM, Jul 01.

7. US Patent and Trademark Office (USPTO) site, http://www.uspto.gov/